\documentclass[11pt]{article}
\usepackage{jcappub}
\usepackage{amsmath}
\usepackage{amssymb}
\usepackage{accents}
\usepackage{subfigure}
\usepackage{color}

\def\dual#1{\accentset{\boldsymbol{\neg}\vspace{-0.2ex}}{#1}}

\newcommand{\nc}{\newcommand}
\nc{\be}{\begin{equation}}
\nc{\ee}{\end{equation}}
\nc{\bea}{\begin{eqnarray}}
\nc{\eea}{\end{eqnarray}}
\nc{\lsp}{\left(}
\nc{\rsp}{\right)}
\nc{\lsc}{\left{}
\nc{\rsc}{\right}}
\nc{\lsb}{\left[}
\nc{\rsb}{\right]}

\title{Attractor behaviour in ELKO cosmology}

\author[a,1]{Abhishek Basak,\note{Corresponding author.}}
\author[a]{Jitesh R. Bhatt}
\author[b]{S. Shankaranarayanan}
\author[b]{and K. V. Prasantha Varma}

\affiliation[a]{Theoretical Physics Group, Physical Research Laboratory, Ahmedabad, India}
\affiliation[b]{Indian Institute of Science Education and Research-Thiruvananthapuram (IISER-TVM), Trivandrum 695016, India}

\emailAdd{abhishek@prl.res.in}
\emailAdd{jeet@prl.res.in}
\emailAdd{shanki@iisertvm.ac.in}
\emailAdd{varma@iisertvm.ac.in}

\abstract{We study the dynamics of ELKO in the context of accelerated phase of 
our universe. To avoid the fine tuning problem associated with the initial conditions, 
it is required that the dynamical equations lead to an early-time attractor. In the 
earlier works, it was shown that the dynamical equations containing ELKO fields do not lead 
to early-time stable fixed points. In this work, using redefinition of variables, 
we show that ELKO cosmology admits early-time stable fixed points. More interestingly, we 
show that ELKO cosmology admit two sets of attractor points corresponding to slow and fast-roll 
inflation. The fast-roll inflation attractor point is unqiue for ELKO as it is independent 
of the form of the potential. We also discuss the plausible choice of interaction terms 
in these two sets of attractor points and constraints on the coupling constant.}

\begin{document}

\maketitle

\section{Introduction}

Inflation is currently one of the successful paradigm of the early
universe\cite{Guth:1980zm,Linde:2005ht}.
The success of inflation not only rests on solving the problems
of the Friedman-Robertson-Walker model, and that it generates 
the primordial spectra of scalar (density) and tensor (gravitational 
waves) perturbations. The temperature variations of CMB as measured in 
WMAP and PLANCK --- to a large extent --- confirm that the these primordial 
density perturbations are generated due to rapid expansion of the 
quantum fluctuations in the early universe. 

Current CMB measurements can at-most provide two physical quantities during inflation: 
(i) inflaton (dynamical field that dominates during inflation) potential and (ii) 
the first derivative of the potential in the observable scales \cite{1995-Kolb.etal-NPSS}. 
It is still unclear what are the properties of the dynamical field that drive inflation. It is 
usually assumed that the field that dominates is a fundamental scalar field. 
While it is the simplest, recently, there has been a surge of activity to look 
at the plausibility that the inflaton has an internal structure \cite{2008-Golovnev.etal-JCAP,Barrow:2009gx,Golovnev:2011yc}. 
One of the key results of the spinor condensate models compared to the standard inflationary 
models is the prediction of running of spectral index in the slow-roll limit that is 
consistent with the current CMB measurements \cite{Gredat:2008qf}.

Even if the observations provide us information about the nature of the dynamical field 
that drive inflation, the initial state of the field will never be known \cite{1997-Lidsey.etal-RMP,Bassett:2005xm,2009-Malik.Wands-PRep}. Einstein's 
equations are non-linear, hence, it is important to know what range of initial conditions 
of inflation can plausibly lead to inflation. For instance, it is possible that if the 
initial velocities of the background field are large then this will stop inflation \cite{1998-Copeland.etal-PRD}. 

The knowledge of initial conditions can provide crucial information about the nature 
of the fields and their interactions with the known matter fields \cite{1998-Copeland.etal-PRD,2008-Copeland.etal-PRD}. 
For instance, it is usually assumed that the inflaton is a fundamental scalar field. However, we do not 
know the nature of the scalar fields or its interaction with other fields. Similarly, 
it is not clear what are the properties of ELKOs and how they interact with the other 
fields. If the observations do provide evidence that the inflation occured due to the 
one of these fields, the initial conditions of these fields will provide information 
about the nature of interactions with standard model particles. This in turn can be useful
for model building which can be verified in high-energy experiments.

The above issues are relevant and imperative to the current acceleration of the universe. 
It is unclear what dynamical fields drive the current accelerating universe. Even if the 
observations reveal the nature of the field, as in the case of inflation, it is not 
possible to know the initial condition that lead to current accelearation. This is referred 
to as cosmic coincidence problem. The constraints on the interaction of these fields with 
standard model particles will provide information about the initial condition that lead to 
acceleration. 

In this work we investigate the following questions: If dynamical field during inflation
is a condensate of the non-standard spinor whether a large set of initial conditions lead 
to inflation. If it does, then, can it constraint the interactions between the spinor 
fields and the standard model particles. Recently a spin-half fermion
called ELKO was proposed as a candidate of dark matter
\cite{Ahluwalia:2004ab,Ahluwalia:2004sz}. Recently. in the literature several authors
have shown that these can lead to accelerated expansion 
\cite{Boehmer:2006qq,Boehmer:2007dh,Boehmer:2010ma,Gredat:2008qf,Shankaranarayanan:2009sz,Shankaranarayanan:2010st,Basak:2011wp}.

In Ref. \cite{Wei:2010ad,Sadjadi:2011uu}, the authors could not find any stable fixed point with 
various kind of potentials. One of the draw-backs of their analyses is the choice of variables. Specifically, 
they have assumed $\phi$ and $V(\phi)$ to be independent variables. In this work, we make a combination 
of $\dot{\phi}$, $H$ and $V(\phi)$ and show that in these newly defined variables the dynamical equations 
have stable fixed points for a wide class of potentials and interactions between ELKO and matter. 
More interestingly, we 
show that ELKO cosmology admit two sets of attractor points corresponding to slow and fast-roll 
inflation. We also discuss the constraints on the coupling constants that will lead to
early-time attractor behavior. 

The manuscript is organised as follows: In sections \ref{sec2} and \ref{sec3} 
we define the background equations and slow-roll parameters.
In section \ref{sec4} we define new sets of variables and rewrite the background equations in terms of 
these variables. A general analysis of fixed point behavior is presented in section \ref{sec5} and 
section \ref{sec6} contains the fixed point analysis for specific potentials. Finally, 
in section \ref{sec7}, we summarize our results and also discuss its implications. 


\section{Background equations\label{sec2}}
The new class of spinors named Elko were proposed as a candidate of dark
matter in Ref.\cite{Ahluwalia:2004ab,Ahluwalia:2004sz}. These
spinors are eigen spinors of charge conjugation operator. These are
non standard spinors(NSS) because unlike the standard spinors Elko spinors have mass
dimension one and $\lsp CPT\rsp^{2}=-\mathbb{I}$. As a consequence these spinors
follow Klein-Gordon equation. Elko spinors are also called `dark' spinor as
its dominant interaction with standard model particles is via Higgs and
gravity. Apart from being considered as inflaton these spinors have drawn lots of 
interests in
works of different authors\cite{daRocha:2008we,HoffdaSilva:2009is,Fabbri:2009aj,Fabbri:2009ka,Fabbri:2010ws}. The
NSS($\lambda$) and its dual($\dual\lambda$) can be defined as 
\be
\lambda=\phi(t)\xi, \qquad
\dual\lambda=\phi(t)\dual\xi,
\label{eq:elko}
\ee
where $\xi$ and $\dual\xi$ are constant spinors with the property
$\dual\xi\xi=\mathbb{I}$, and $\phi(t)$ is a scalar which is time depepndent. The
action can be written as [boeh]
\be
S=\int
d^{4}x\sqrt{-g}\lsb\frac{1}{2}g^{\mu\nu}\dual{\lambda}\overleftarrow{\nabla}_{(\mu}\nabla_{\nu)}\lambda-V\lsp\dual\lambda\lambda\rsp\rsb,
\label{eq:action}
\ee
where $\nabla_{\mu}$ is the covariant derivative and the round bracket in
the subscript denotes symmetrisation. Varying the action (\ref{eq:action})  
the energy density and pressure can be written in terms as 
\bea
\nonumber\rho_{\phi}&=&\frac{1}{2}\dot\phi^{2}+\frac{3}{8}H^{2}\phi^{2}+V(\phi),\\
p_{\phi}&=&\frac{1}{2}\dot\phi^{2}-\frac{1}{4}\left(H\phi^{2}\right)^{.}-
              \frac{3}{8}H^{2}\phi^{2}-V(\phi),
\label{eq:ep1}
\eea
where $H$ is the Hubble parameter.
The equation of motion can be written as:
\be
\ddot\phi+3H\dot\phi-\frac{3}{4}H^{2}\phi+V_{,\phi}=0,
\label{eom}
\ee

\section{Slow-roll parameters for ELKOs\label{sec3}}

Due to the presence of the $H^2 \, \phi^2$ term in the density and pressure, 
one has be careful in defining the slow-roll parameters for the ELKO condensate.
In this section, we give the expressions for the slow-roll parameters. From 
the expression of time-time component of
Einstein's equation one can write the Friedman's equation for NSS as:
\be
H^2 = \frac{1}{3M_{pl}^2}\lsp\frac{\dot{\varphi}^2}{2D}+\hat{V}\rsp
\label{fr1}
\ee
where $D=1-\tilde{F} =1-\frac{\varphi^2}{8M_{pl}^2}$ and
$\hat{V}=\frac{V}{D}$. Taking the time derivative on the both the sides
of equation (\ref{fr1}) one can write the slow-roll parameter $\epsilon$ as
\be
\epsilon=-\frac{\dot
H}{H^{2}}=\frac{3}{2}\frac{\dot{\phi}^{2}/D}{\dot{\phi}^{2}/2D+\hat{V}}+\frac{\dot
D}{HD}=\epsilon_{\rm{can}}+\alpha.
\label{epsilon}
\ee
where $\alpha=\frac{\dot D}{HD}\ll1$ is an extra parameter which appears in the definition 
of the slow-roll parameter for ELKOs and $\epsilon_{\rm{can}}$ is the slow-roll parameter defined 
for the canonical single scalar field,
$\epsilon_{\rm{can}}=\frac{3}{2}\frac{\dot{\phi}^{2}}{\dot{\phi}^{2}/2+V}\simeq\frac{3}{2}\frac{\dot{\phi}^{2}}{V}$
(when $\dot{\phi}^{2}\ll V$). 
\par
Substituting $\lsp\dot{\phi}^{2}/2D+\tilde{V}\rsp$ from (\ref{epsilon}) in
(\ref{fr1}) one can write 
\be
\dot{\phi}^{2}=2M_{pl}^{2}H^{2}D\lsp\epsilon-\alpha\rsp.
\label{del0}
\ee
Now in this case we define $\delta=\frac{\ddot\phi}{H\dot\phi}$. Taking the time derivative on both the sides of (\ref{del0}) we get,
\be
\frac{\ddot\phi}{H\dot\phi}=\delta=-\epsilon+\frac{\alpha}{2}+\frac{\lsp\epsilon-\alpha\rsp^{.}}{2H\lsp\epsilon-\alpha\rsp}
\ee
The last term can be dropped as it is the time derivative of the slow-roll
parameters. Therefore finally one can write the definition of $\delta$ as:
\be
\delta=-\epsilon+\frac{\alpha}{2} = - \epsilon_{\rm can} - \frac{\alpha}{2}.
\ee
A closer inspection of above expression immediately suggests that $\delta$ is negative definite.
For canonical scalar field, it is positive definite. This has an important effect for the spectra of 
scalar perturbations \cite{Gredat:2008qf}. 


\section{Dynamical equations\label{sec4}}
The expressions of energy density and pressure (\ref{eom}) can be written in
terms of newly defined variables $X$ and $\tilde{V}$ as 
following
\be
 \rho_{\phi}=X+\tilde{V}, \qquad 
 p_{\phi}=X-\tilde{V},
 \label{eq:ep2}
\ee
where 
\begin{eqnarray}
\label{eq:defX}
X & =& \frac{1}{2}\dot\phi^{2}-\frac{1}{8}\left(H\phi^{2}\right)^{.} \\
\label{eq:defV}
\tilde{V}&=&\frac{1}{8}\left(H\phi^{2}\right)^{.}+\frac{3}{8}H^{2}\phi^{2}+V(\phi) \, .
\end{eqnarray}
Physically, $X$ refers to the kinetic energy of the condensate field and $\tilde{V}$ refers 
to the potential energy of the condensate. Note that $H \dot{\phi} \phi$ acts like a friction 
term while $\dot{H} \phi^2$ acts as an anti-friction term. Elko dynamics crucial depends on 
which of these two terms dominate during inflation. 

Friedmann equation can be written as 
\begin{equation}
 H^{2}=\frac{\kappa^{2}}{3}\rho_{tot}=\frac{\kappa^{2}}{3}\left(\rho_{\phi}+\rho_{m}\right),
 \label{eq:fr}
\end{equation}
where $\rho_{m}$ is the matter density and $\kappa^{2}=8\pi G$. Using equation (\ref{eq:ep2}) we
can write the Friedmann equation(\ref{eq:fr}) as 
\begin{equation}
x^{2}+y^{2}+v^{2}=1,
\label{eq:cons}
\end{equation}
where $x$,$y$ and $v$ can be defined as $x=\frac{\kappa\sqrt{X}}{\sqrt{3}H}$,
$y=\frac{\kappa\sqrt{\tilde{V}}}{\sqrt{3}H}$ and $v=\frac{\kappa\sqrt{\rho_{m}}}{\sqrt{3}H}$. 
Now, if we consider that the matter and dark energy are interacting only with themselves then
the continuity equation 
\begin{equation}
 \dot\rho_{\rm tot}+3H(\rho_{\rm tot}+p_{\rm tot})=0
\end{equation}
can be written as two separate equations
\begin{equation}
 \dot\rho_{\phi}+3H(\rho_{\phi}+p_{\phi})=-Q,
 \label{eq:ce1}
\end{equation}
\begin{equation}
\dot\rho_{\rm m}+3H(\rho_{\rm m}+p_{\rm m})=Q,
\label{eq:ce2}
\end{equation}
where $Q$ is the interaction term.
In terms of the variables $x$, $y$, $v$ equations (\ref{eq:ce1},\ref{eq:ce2}) can be written respectively as
\begin{equation}
 x^{\prime}=\left(\epsilon-3\right)x-\frac{\lambda}{H}\frac{y^{2}}{x}-\frac{Q_{1}}{x},
 \label{eq:de1}
\end{equation}
\begin{equation}
 v^{\prime}=\left(\epsilon-\frac{3}{2}\gamma\right)v+\frac{Q_{1}}{v}.
 \label{eq:de2}
\end{equation}
Here $^{\prime}$ is the derivative with respect to time divided by $H$,
$\epsilon=-\frac{\dot H}{H^{2}}$ 
and $\lambda=\frac{\dot{\tilde{V}}}{\tilde{V}}$, $Q_{1}=\frac{\kappa^{2}Q}{6H^{3}}$. 
To derive the above equations we have used the relation
$p_{m}=\left(\gamma-1\right)\rho_{m}$, where $\gamma$ takes
the value either $1$ or $\frac{4}{3}$ depending on whether the universe is
filled with cold matter or radiation respectively.
Derivative of the variable $y$ with respect to time give us
\begin{equation}
 y^{\prime}=\left(\epsilon+\frac{\lambda}{2H}\right)y.
  \label{eq:de3}
  \end{equation}
$\dot H$ can be written as 
\begin{eqnarray*}
\dot H=-\frac{\kappa^{2}}{2}\left[\rho_{\phi}+p_{\phi}+\rho_{m}+p_{m}\right]
\label{eq:rc}
\end{eqnarray*}
Therefore we have three dynamical equations (\ref{eq:de1}), (\ref{eq:de2}) and
(\ref{eq:de3}) with one constraint (\ref{eq:cons}). It is important to contrast 
the above set of variables with those used earlier \cite{Wei:2010ad}. The two 
variables $X$ and $\tilde{V}$ are independent of each other. However, in Wei's 
analysis \cite{Wei:2010ad}, the two variables $y$ and $u$ are not independent. 

In the rest of this work, we study the stability of fixed points with these equations. 
We show that ELKOs show a new set of fixed points that can not be identified 
directly with the canonical scalar field. 


\section{Fixed points and stability analysis: General Analysis\label{sec5}}

Fixed points are those points where the dynamical variables stop evolving,
i.e., at fixed point $(\bar{x},\bar{y},\bar{v})$ the time derivative of
$x$, $y$ and $v$ are zero. At fixed points, dynamical equations (\ref{eq:de1}, 
\ref{eq:de2}, \ref{eq:de3}) can be written as:
\be
\left(\bar{\epsilon}-3\right)\bar{x}-\frac{\lambda}{H}\frac{\bar{y}^{2}}{\bar{x}}-\frac{Q_{1}}{\bar{x}}=0,
\label{eq:de11}
\ee
\be
\left(\bar{\epsilon}-\frac{3}{2}\gamma\right)\bar{v}+\frac{Q_{1}}{\bar{v}}=0,
\label{eq:de12}
\ee
\be
\left(\bar{\epsilon}+\frac{\lambda}{2H}\right)\bar{y}=0.
\label{eq:de13}
\ee
Eq. (\ref{eq:de13}) leads to two set of fixed points: 
\begin{enumerate}
\item Case I: $\bar{y}=0$ and $\bar{\epsilon} \neq - \frac{\lambda}{2H}$
\item Case II: $\bar{y} \neq 0$ and $\bar{\epsilon} = - \frac{\lambda}{2H}$
\end{enumerate}
In the rest of this section, we will consider the above two cases 
for general interaction term $Q_1$. In the following section, we 
consider special cases for the interaction term and discuss 
the nature of fixed points. 

\subsection{Case I}

Substituting this value of $\bar{y}$ in equation  (\ref{eq:de11}) we get 
\begin{equation}
\bar{\epsilon} = 3 \bar{x}^{2} + \frac{Q_{1}}{\bar{x}^2}
\label{eq:caseIepsilon}
\end{equation}
The above form of $\epsilon$ gives crucial information about the 
class of interaction terms between the condensate and matter fields 
that can lead to attractor behavior. In particular, it immediately 
shows that $Q_{1} \propto x^2$ may not lead to stable attractor points. 
Also it provides an upper bound on the coupling constant. We discuss 
these in the next section. 

General expression for $\epsilon$ can be written as
\begin{equation}
 \epsilon=-\frac{\dot H}{H^{2}}
  =\frac{3}{2}\gamma+\left(3-\frac{3}{2}\gamma\right)x^{2}-\frac{3}{2}\gamma y^{2}.
  \label{eq:epsilon}
\end{equation}
Therefore at fixed points $\bar{\epsilon}$ can be written as 
$\bar{\epsilon}=\frac{3}{2}\gamma+\left(3-\frac{3}{2}\gamma\right)\bar{x}^{2}$.
$\bar{\epsilon}$ 
is a positive quantity ensuring accelerated expansion of the universe.
Finally one can write an important relation for $\bar{\epsilon}$ which will be used later 
\be
\bar{\epsilon}-3=\lsp\frac{3}{2}\gamma-3\rsp\lsp1-\bar{x}^{2}\rsp.
\label{bars}
\ee
\par
Once we get the fixed points, we need to study the stability of the fixed point to ensure that
the fixed points are actually giving us an attractor. If the fixed points
are stable then we can have the attractor and finally we will be  
able to alleviate the `cosmic coincidence' problem. To analyse the
stability of these fixed points we perturb 
the system about the fixed point, $x\rightarrow \bar{x}+\delta x$ and 
$y\rightarrow\bar{y}+\delta y$. Then we study the evolution of the
perturbations. If we have a growing solution of the perturbations our fixed
points are not stable, however if we find a decaying solution we can say
that our fixed points are stable. Substituting these values of $x$ and $y$
in equation (\ref{eq:de1}) and 
(\ref{eq:de3}) we get the perturbed equations of $x$ and $y$ as follows:
\bea
\nonumber\delta
x^{\prime}&=&\lsb\lsp\bar{\epsilon}-3\rsp+\lsp6-3\gamma\rsp\bar{x}^{2}+
                                     \frac{Q_{1}}{\bar{x}^{2}}-
                                     \frac{1}{\bar{x}}\frac{\partial
                                     Q_{1}}{\partial x}\rsb\delta x-\\
                               &&\lsp\frac{1}{\bar{x}}\frac{\partial
                               Q_{1}}{\partial y}\rsp\delta y,
\label{eq:pe1}                               
\eea
\begin{equation}
 \delta y^{\prime}=\left(\bar{\epsilon}+\frac{\lambda}{2H}\right)\delta y.
 \label{eq:pe2}
\end{equation}
Here we have used $\delta\epsilon=\left[\left(6-3\gamma\right)\bar{x}\right]\delta x$ and 
$\bar{y}=0$. Equations (\ref{eq:pe1}) and (\ref{eq:pe2}) can be written as 
\begin{equation}
 \begin{pmatrix}
   \delta x^{\prime}\\
   \delta y^{\prime}
 \end{pmatrix}=
 \left(M\right)
 \begin{pmatrix}
   \delta x\\
   \delta y            
 \end{pmatrix},
\end{equation}
where $$M=\begin{pmatrix}
   \left(\bar{\epsilon}-3\right)+\left(6-3\gamma\right)\bar{x}^{2}+\frac{Q_{1}}{\bar{x}^{2}}-
   \frac{1}{\bar{x}}\frac{\partial Q_{1}}{\partial x} & 
   \frac{1}{\bar{x}}\frac{\partial Q_{1}}{\partial y}\\
   0 & \left(\bar{\epsilon}+\frac{\lambda}{2H}\right)
 \end{pmatrix}.$$
Two eigenvalues of the matrix $M$ are
\begin{equation}
 \mu_{1}=\left(\bar{\epsilon}+\frac{\lambda}{2H}\right),
\end{equation}
\begin{equation}
 \mu_{2}=\left(\bar{\epsilon}-3\right)+\left(6-3\gamma\right)\bar{x}^{2}+\frac{Q_{1}}{\bar{x}^{2}}-
                \frac{1}{\bar{x}}\frac{\partial Q_{1}}{\partial x}.
\end{equation}
Stability around the fixed points depend upon the nature of the eigen
values $\mu_{1}$ and $\mu_{2}$. When $\mu_{1}<0$, $\mu_{2}<0$ the fixed
points are stable and we can get an attractor solution. If $\mu_{1}>0$,
$\mu_{2}>0$, the fixed points are unstable and we can not have any
attractor. If one of them is
positive and other one is negative, we get a saddle point which says that
at one direction the fixed points are stable and at the other direction the
fixed points are unstable. 

In Ref.\cite{Wei:2010ad} it was noted that equation-of-state parameter
$w_{\phi}=\frac{p_{\phi}}{\rho_{\phi}}\geq-1$ when
$\dot{\phi}^{2}\geq\frac{1}{4}\lsp H\phi^{2}\rsp^{.}$. The dark energy will
enter the phantom region ($w_{\phi}<-1$) if
$\dot{\phi}^{2}\leq\frac{1}{4}\lsp H\phi^{2}\rsp^{.}$. Therefore in the
region $w_{\phi}\geq-1$ we always get $X>0$. Now from Freidmann equation (\ref{eq:fr}) we get
\begin{equation}
H^{2}=\frac{8\pi G}{3}\left(X+\tilde{V}+\rho_{m}\right),
\end{equation}
which implies that $H^{2}>\frac{8\pi G}{3}\tilde{V}$. Finally taking logarithmic time
derivative on both the sides of this inequality we get
\begin{equation}
\epsilon+\frac{\lambda}{2H}<0.
\end{equation}
This means that $\mu_{1}$ is always negative as far as $w_{\phi}\geq-1$ is
concerned. Therefore it is possible to have a stable fixed point if
$\mu_{2}$ becomes negative for some interaction $Q$. 
In the next section, we analyse this for three types of interactions.

\subsection{Case II}

Substituting the value of $\bar{\epsilon}$ and constraint (\ref{eq:cons}) in Eqs. (\ref{eq:de1}) and (\ref{eq:de2}), 
we get for ($\gamma = 1$):
\begin{eqnarray}
\label{eq:Case2-Const1}
x^{'} &=& (\bar{\epsilon} - 3) x - \frac{\lambda}{H}\frac{y^2}{x} - \frac{Q_1}{x} \\
\label{eq:Case2-Const2}
v^{'} &=& \left(\bar{\epsilon} - \frac{3}{2}\right) v + \frac{Q_1}{v} 
\end{eqnarray}
Substituting for $\lambda$, we get,
\begin{eqnarray}
\label{eq:case2epsilon}
\bar{\epsilon} &=& - \frac{\lambda}{2H} = 3x^2 + \frac{3}{2}v^2 \\
\delta \epsilon &=& 6x \delta x + 3v \delta v
\end{eqnarray}
The perturbed equations about the fixed point are:
\begin{eqnarray}
\label{eq:case2pert1}
 \delta x^{'} &=& \left( 3 - 9x^2 - \frac{15}{2}v^2 -3\frac{v^2}{x^2} + 3\frac{v^4}{x^2} \right) \delta x + 
 \left( 6\frac{v}{x} - 12\frac{v^3}{x} - 15 xv \right) \delta v - \delta (\frac{Q_1}{x}) \\
 \label{eq:case2pert2}
\delta v^{'} &=& 6 \, x \, v \, \delta x + \left( 3x^2 + \frac{9}{2}v^2 - \frac{3}{2} \right) \, \delta v + 
\delta \left( \frac{Q_1}{v}\right)
\end{eqnarray}
These attractor points are unique to ELKO cosmology regarding which we would like to stress the following 
points:
\begin{enumerate}
\item The perturbed equations do not explicitly depend on the potential. Hence, these equations can 
 be realised for any potential provided $\bar{\varepsilon} = - \lambda/(2 H)$ is satified. 
\item If $x \ll 1$ and $v \to 1$ (or vice-versa), Eq. (\ref{eq:case2epsilon}) implies that $\epsilon > 1$. 
It is easy to see that $\ddot{a}(t) > 0$ implying that this corresponds to fast-roll inflation \cite{2005-Starobinsky-JETPL}. 
\end{enumerate}

%

\section{Special cases of the interaction term\label{sec6}}

In the previous section, we have obtained the 
condition for the existence of fixed point 
for general interaction. However, the analysis 
for a general interaction term is complicated. 
Here, for two cases, we take simple form of the 
interaction term and show explicitly the nature 
of the fixed points. 

\subsection{Case I: Slow-roll}

\subsubsection{\Large{$Q_{1}= \beta v^{2}x$}}
In this case the fixed point $\bar{y}$ is zero. The fixed point $\bar{x}$
and $\bar{v}$ can be found using the equation (\ref{eq:de11}).
Substituting $\bar{y}=0$ in equation (\ref{eq:de11}) and using
$\bar{v}^{2}=1-\bar{x}^{2}$ from equation (\ref{eq:cons}) we can write 
\begin{equation}
 (\bar{\epsilon}-3)\bar{x}-\frac{Q_{1}}{\bar{x}}=0.
 \label{fp}
\end{equation}
Which gives us two solutions for $\bar{x}$
\be
\bar{x}=\pm 1, \qquad 
\bar{x}=-\frac{\beta}{\left(3-\frac{3}{2}\gamma\right)}.
\label{barx1}
\ee
Now $\bar{x}=\pm 1$ can not be a scaling solution because that will make our universe completely
dark-energy dominated. Therefore the only possible solution is 
$\bar{x}=-\frac{\beta}{\left(3-\frac{3}{2}\gamma\right)}$ which is negative as $\beta$ and
$\left(3-\frac{3}{2}\gamma\right)$ are both positive.
\par
Therefore in this case the eigenvalue $\mu_{2}$ of the matrix $M$ can be written as
\begin{equation}
 \mu_{2}=\left[-\left(3-\frac{3}{2}\gamma\right)\lsp1-3\bar{x}^{2}\rsp+2\beta\bar{x}\right].
\label{mu21}                                     
\end{equation}
Now substituting the solution of $\bar{x}$ from (\ref{barx1}) in the above
expression of $\mu_{2}$ one can get the following expression of $\mu_{2}$:
\be
\mu_{2}=-\lsp3-\frac{3}{2}\gamma\rsp+\frac{\beta^{2}}{\lsp3-\frac{3}{2}\gamma\rsp},
\label{mu22}
\ee
From the above expression of $\mu_{2}$ it can be understood that when the
first term dominates over the last term one can get $\mu_{2}<0$. Therefore the
condition for having a stable fixed point for this kind of interaction is:
\be
\beta<\lsp3-\frac{3}{2}\gamma\rsp.
\ee
Therefore in this kind of interaction the coupling constant can not be very
large and $\bar{\epsilon}$ is always less than 1.


\subsubsection{{$Q_{1}=\beta v^{2}x^{2}$}} In this case
equation (\ref{bars}) and (\ref{fp}) tell us that
\be
\lsp\frac{3}{2}\gamma-3\rsp\lsp1-\bar{x}^{2}\rsp\bar{x}-\beta\lsp1-\bar{x}^{2}\rsp\bar{x}=0.
\ee
So, the only solution of  $\bar{x}=\lsp0,\pm 1\rsp$. One can not take these
solutions as fixed points as the universe will become
purely matter dominated and dark-energy dominated respectively in those
cases. Therefore we do not have any physical fixed points in this kind of
interaction.  

\subsubsection{{$Q_{1}=\beta v x^{2}$}}
Following the similar method as described above using 
(\ref{fp}) for this kind of interaction one can find that at fixed
point the only solution for $x$ is:
\be
\bar{x}=\pm\sqrt{1-\frac{\beta^{2}}{\lsp3-\frac{3}{2}\gamma\rsp^{2}}},
\label{barx2}
\ee
Here we have considered $\bar x\neq\lsp0,\pm1\rsp$. 
Substituting the above expression of $Q_{1}$ in the expression of $\mu_{2}$
one can get 
\be
\mu_{2}=\lsp6-3\gamma\rsp\bar{x}^{2}+\beta\frac{\bar{x}^{2}}{\bar v}.
\ee
Using the definition of $\bar{v}^{2}=1-\bar{x}^{2}$ and the expression of
$\bar x$ from (\ref{barx2}) one can write the expression of $\mu_{2}$ in
terms of the coupling $\beta$ as:
\be
\mu_{2}=\lsp3-\frac{3}{2}\gamma\rsp-\frac{\beta^{2}}{\lsp3-\frac{3}{2}\gamma\rsp},
\label{mu23}
\ee
Therefore in this case the $\mu_{2}$ will be negative only when
$\beta>\lsp3-\frac{3}{2}\gamma\rsp.$ However from (\ref{barx2}) one can
see that this condition will make $\bar x$ imaginary. Therefore we can not
find a stable fixed point in this case.

\subsection{Case II: Fast roll}

\subsubsection{\Large{$Q_{1}=\beta v^{2}x$}}

For this interaction, the perturbed equations of $x$ and $v$ are:
\begin{eqnarray}
\label{eq:case2Ipert1}
\delta x^{'} &=&  [3 - 9x^2 - \frac{15}{2}v^2 -3\frac{v^2}{x^2} + 
3\frac{v^4}{x^2}]\delta x + [6\frac{v}{x} - 12\frac{v^3}{x} - 15 xv - 2 \beta v ]\delta v.\\
\label{eq:case2Ipert2}
\delta v^{'} &=& [6 x v  + \beta v] \delta x + [3x^2 + \frac{9}{2}v^2 - \frac{3}{2} + \beta x]\delta v 
\end{eqnarray}
The two eigen-values corresponding to the above set of equations are negative. Fig. (1a) shows that for 
different initial conditions $v \to 1$ and $x \ll 1$ is an attractor point. 

\subsubsection{\Large{$Q_{1}=\beta v^{2}x^{2}$}}

For this interaction, the perturbed equations of $x$ and $v$ are:
\begin{eqnarray}
\label{eq:Case2IIpert1}
\delta x^{'} &=&  [3 - 9x^2 - \frac{15}{2}v^2 
-3\frac{v^2}{x^2} + 3\frac{v^4}{x^2} - \beta v^2]\delta x 
+ [6\frac{v}{x} - 12\frac{v^3}{x} - 15 xv - 2 \beta v x]\delta v.\\
\label{eq:Case2IIper2}
\delta v^{'} &=& [6 x v  + 2\beta v x] \delta x + [3x^2 + \frac{9}{2}v^2 - \frac{3}{2} + \beta x^2]\delta v 
\end{eqnarray}
Here again, both the eigenvalues corresponding to the above set of equations are negative. 
The eigenvalues are negative for all ranges of $\beta$ for which $x$ and $v$ are real.
Fig. (1b) shows that for different initial conditions $v \to 1$ and $x \ll 1$ 
is an attractor point. 

\subsubsection{\Large{$Q_{1}=\beta \, v \, x^{2}$}}

For this interaction, the perturbed equations of $x$ and $v$ are:
\begin{eqnarray}
\label{eq:Case2IIIPert1}
\delta x^{'} &=&  [3 - 9x^2 - \frac{15}{2}v^2 
-3\frac{v^2}{x^2} + 3\frac{v^4}{x^2} - \beta v]\delta x 
+ [6\frac{v}{x} - 12\frac{v^3}{x} - 15 xv -  \beta x]\delta v.\\
\label{eq:Case2IIIPert2}
\delta v^{'} &=& [6 x v  + 2 \beta x] \delta x + [3x^2 + \frac{9}{2}v^2 - \frac{3}{2}]\delta v 
\end{eqnarray}

Here again, both the eigenvalues corresponding to the above set of equations are negative for 
all valus of $\beta$ where $x$ and $v$ are real.
Fig. (1c) shows that for different initial conditions $v \to 1$ and $x \ll 1$ 
is an attractor point. 

\section{Conclusion\label{sec7}}

As it is known, it is not possible to know precisely the initial condition 
of the field that drives inflation. Hence, any model of inflation need to 
show that attractor points exists in the space of the matter field variables. 
In all previous work, it was not possible to show explicitly that the Elko 
condensate does indeed lead to late-time attractor behavior. 

In this work we have shown that rewritting the background field equations interms of 
new variables $X$ and $\tilde{V}$, one could show the existence of late-time 
attractor. Interestingly, we notice that the two set of attractor points exist for 
ELKO condensates. In case I, which is similar to canonical scalar field inflation 
\cite{1998-Copeland.etal-PRD}, the attractors are realised only when $\bar{\epsilon} < 1$.
In case II, which is unique for ELKO cosmology, the attractors are realised only 
when $\bar{\epsilon} > 1$ and they exist independent of the form of the potential. 
It will be interesting to repeat Starobinsky's analysis \cite{2005-Starobinsky-JETPL}
for this case and see whether the spectrum of perturbations are nearly scale-invariant.

In the phantom region $\lsp w_{\phi}\leq-1\rsp$ obtaining stable fixed point may not be
possible as in that case $X<0$ may not allow $H^{2}>\tilde{V}$, in other
words in phantom region $\mu_{1}<0$ may not be possible always. 
In this work the variables we have worked with are not a simple transformation 
of the variables chosen by Ref.\cite{Wei:2010ad}. The stability may be achieved 
by redefining the potential and the kinetic part. This is currently work under progress.

\section{Acknowledments}

SS acknowledges the support of DST, Government of India through Ramanujan
fellowship and Max Planck-India Partner Group on Gravity and
Cosmology. KVPV is supported by DST, Goverment of India through KVPY fellowship.

\providecommand{\href}[2]{#2}\begingroup\raggedright\endgroup

\end{document}